\documentclass{article}

\pdfoutput=1 

 \usepackage[final, nonatbib]{neurips_2020}

\usepackage[utf8]{inputenc} 
\usepackage[T1]{fontenc}    
\usepackage{hyperref}       
\usepackage{url}            
\usepackage{booktabs}       
\usepackage{amsfonts}       
\usepackage{nicefrac}       
\usepackage{microtype}      
\usepackage{xcolor}
\usepackage{ulem}
\usepackage{graphicx}
\usepackage{subcaption}
\usepackage{amsmath}
\usepackage{bbold}

\graphicspath{
    {imgs/}
}

\title{Towards disease-aware image editing of chest X-rays}

\author{

   \hspace*{0.5cm}Aakash Saboo\\
  \hspace*{0.5cm}Caring Research\\
  \hspace*{0.5cm}India\\
  \texttt{\hspace*{0.5cm}aakashsaboo2@gmail.com} 
  \And
  Sai Niranjan Ramachandran\thanks{Work done during an internship at Caring Research, India.}\\
 Indian Institute of Science\\
  India\\
  \texttt{rsainiranjan@iisc.ac.in} \\
  \And
  Kai Dierkes\\
  Pupil Labs\\
  Germany\\
  \texttt{\quad\quad kai@pupil-labs.com \quad} \\
  \And
  \hspace*{-0.5cm}Hacer Yalim Keles\\
  \hspace*{-0.5cm}Ankara University\\
  \hspace*{-0.5cm}Turkey\\
  \hspace*{-0.5cm}\texttt{hkeles@ankara.edu.tr} }
  
\begin{document}

\maketitle

\begin{abstract}
Disease-aware image editing by means of generative adversarial networks (GANs) constitutes a promising avenue for advancing the use of AI in the healthcare sector. Here, we present a proof of concept of this idea.
While GAN-based techniques have been successful in generating and manipulating natural images, their application to the medical domain, however, is still in its infancy. 
Working with the CheXpert data set, we show that StyleGAN can be trained to generate realistic chest X-rays. 
Inspired by the Cyclic Reverse Generator (CRG) framework, we train an encoder that allows for faithfully inverting the generator on synthetic X-rays and provides organ-level reconstructions of real ones. 
Employing a guided manipulation of latent codes, we confer the medical condition of cardiomegaly (increased heart size) onto real X-rays from healthy patients. 

%
\end{abstract}

\section{Introduction}
In the face of the global COVID-19 pandemic, the development of efficient means for the accurate diagnosis of pulmonary diseases took on a new urgency.
Many image-based classification systems are being developed for the automated diagnosis of COVID-19 on chest X-rays ~\cite{brunese_explainable_2020,panwar_deep_2020}.
While attempting to create a classification and localization system for diagnosing COVID-19 on chest X-rays, we explored an approach that involved disentangling healthy and diseased image features.  
We propose a conceptual framework to address this issue, using cardiomegaly as an example. Cardiomegaly, or increased size of the heart, is a key imaging feature that can distinguish between Acute Respiratory Distress Syndrome (ARDS) due to COVID-19 pneumonia and pulmonary edema due to heart failure~\cite{archer_stephen_l_differentiating_2020, ZhuTan20}.


Generative models such as GANs constitute powerful techniques for capturing the distribution of high-dimensional image data and synthesizing realistic images \cite{KarAil18, KarLai18, KarLai20, VahKau20}.
GANs have been successfully employed for data augmentation \cite{SanYan19} as well as for the manipulation of images in a semantically controlled manner \cite{DogKel20,GyaLi19}.
Most studies to date, however, have been restricted to natural images. 
Due to the relative lack of labelled high-quality image data, transferring existing techniques from natural to medical images, such as chest X-rays, poses its own specific set of obstacles. 


We argue that any suitable AI system for the generation and disease-aware editing of medical images needs to fulfil three criteria. 
(i) It needs to provide a generator $G$, which based on latent codes produces images statistically equivalent to real images. In particular, $G$ needs to be sufficient, i.e. any given real image needs to be generatable from a suitable latent code.
(ii) In order to allow for disease-aware image editing based on the manipulation of latent codes, disease-related characteristics need to be disentangled in latent space. 
(iii) The system needs to provide an encoder $E$ for embedding a given image into latent space, i.e. a means for faithfully inverting the generation process. 

Here, we show that StyleGAN \cite{KarLai18} can be trained on the CheXpert \cite{IrvRaj19} dataset to provide a generator $G$ of chest X-rays at 128x128 resolution. 
Taking inspiration from the recently proposed CRG framework~\cite{DogKel20}, we train a corresponding encoder $E$ and assess the extent to which $G$ and $E$ fulfil the above criteria. 
In particular, we argue that the medical condition cardiomegaly is at least partially disentangled in latent space, thus permitting disease-aware image manipulation. 
Our preliminary results provide a compelling proof of concept and constitute an important step towards the generation and disease-aware editing of chest X-rays.
\begin{figure}[!tbp]
\includegraphics[width=\textwidth]{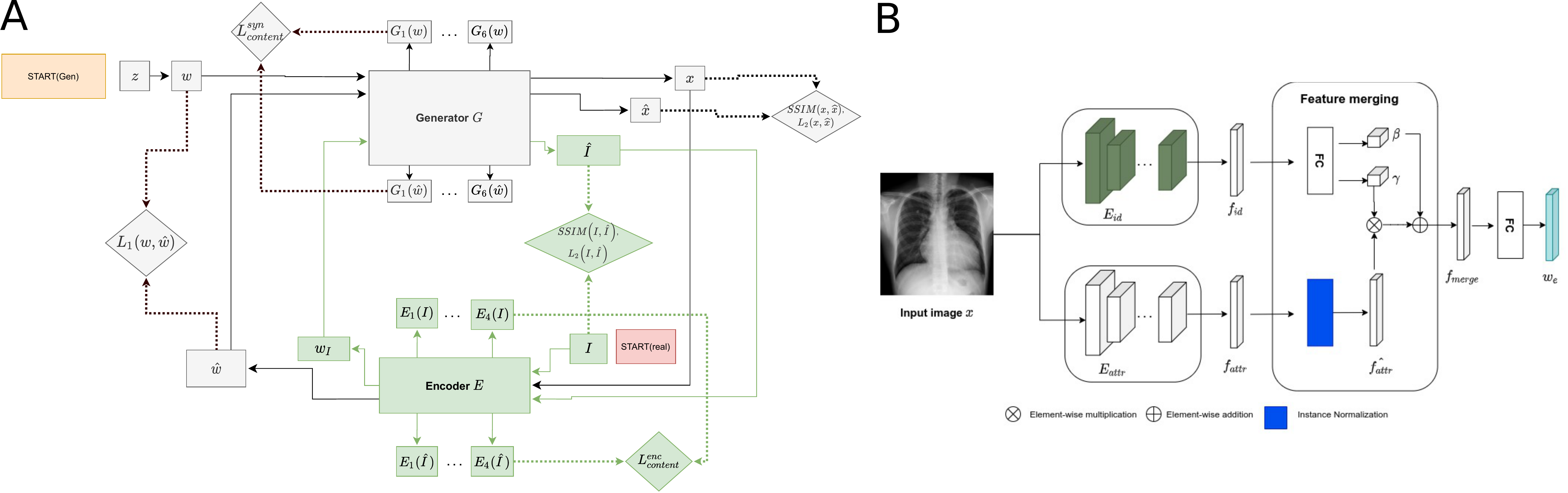}
\caption{(A) CRG architecture; (B) Encoder architecture. Best viewed in electronic version. }
 \label{fig:architectures}
\end{figure}


\section{The Method}
In our CRG setup see Fig.~\ref{fig:architectures}A, we use StyleGAN as generator $G$ and a two branch Resnet-101 based architecture as encoder $E$. We took inspiration from \cite{GuaTai20} in designing the architecture of $E$ and furthermore used denormalization as suggested in \cite{ParWan19} for merging the features from both branches. The high-level architecture of $E$ is depicted in Fig.~\ref{fig:architectures}B. In a first step, we train $G$ using frontal X-rays from the CheXpert dataset in 128x128 pixel resolution. We then train $E$, while keeping the weights of $G$ fixed and furthermore using a constant noise realization. 
During training, the weights of $E$ are updated two times in each iteration as proposed in the original CRG work \cite{DogKel20}. In the first update, following the cyclic path depicted with black lines in Fig.~\ref{fig:architectures}A, we back-propagate a loss $L_{\rm total}^{\rm syn}$ designed to minimize the difference between generated synthetic images and their reconstructions. 
More specifically, $L_{\rm total}^{\rm syn} = L_{1}(w,\hat{w})+ L_{2}(x,\hat{x})+{\rm SSIM}(x,\hat{x})+L_{\rm content}^{\rm syn}$, where $w$-s and $x$-s refer to latent codes and images, respectively, SSIM is the structural similarity metric, and $L_{\rm content}^{\rm syn}=\sum_{i=0}^{L}\frac{1}{N_i}\left\|\it{G}_{i}(w)-\it{G}_{i}(\hat{w})\right\|_{1}$. Here, $\it{G}_{i}$ denotes the features from the $i^{th}$ hidden layer of $G$, $N_i$ is the total size of the corresponding feature map, and $L$ is the number of hidden layers in the network.  
The second update is designed to optimize the quality of real image reconstructions and follows the loop shown by green lines in Fig.~\ref{fig:architectures}A. 
Using an analogous notation as above, the loss back-propagated in this cycle is defined as $L_{\rm total}^{\rm real} = L_{2}(I,\hat{I})+{\rm SSIM}(I,\hat{I})+L_{\rm content}^{\rm enc}$, where $L_{\rm content}^{\rm enc}=\sum_{i=0}^{L}\frac{1}{N_i}\|E_{i}(I)-E_{i}(\hat{I})\|_{1}$. Note, however, this time content loss is computed using the features of the encoder.

\section{Results and Discussions}
\begin{figure}[t!]
     \centering
      \includegraphics[width=\textwidth]{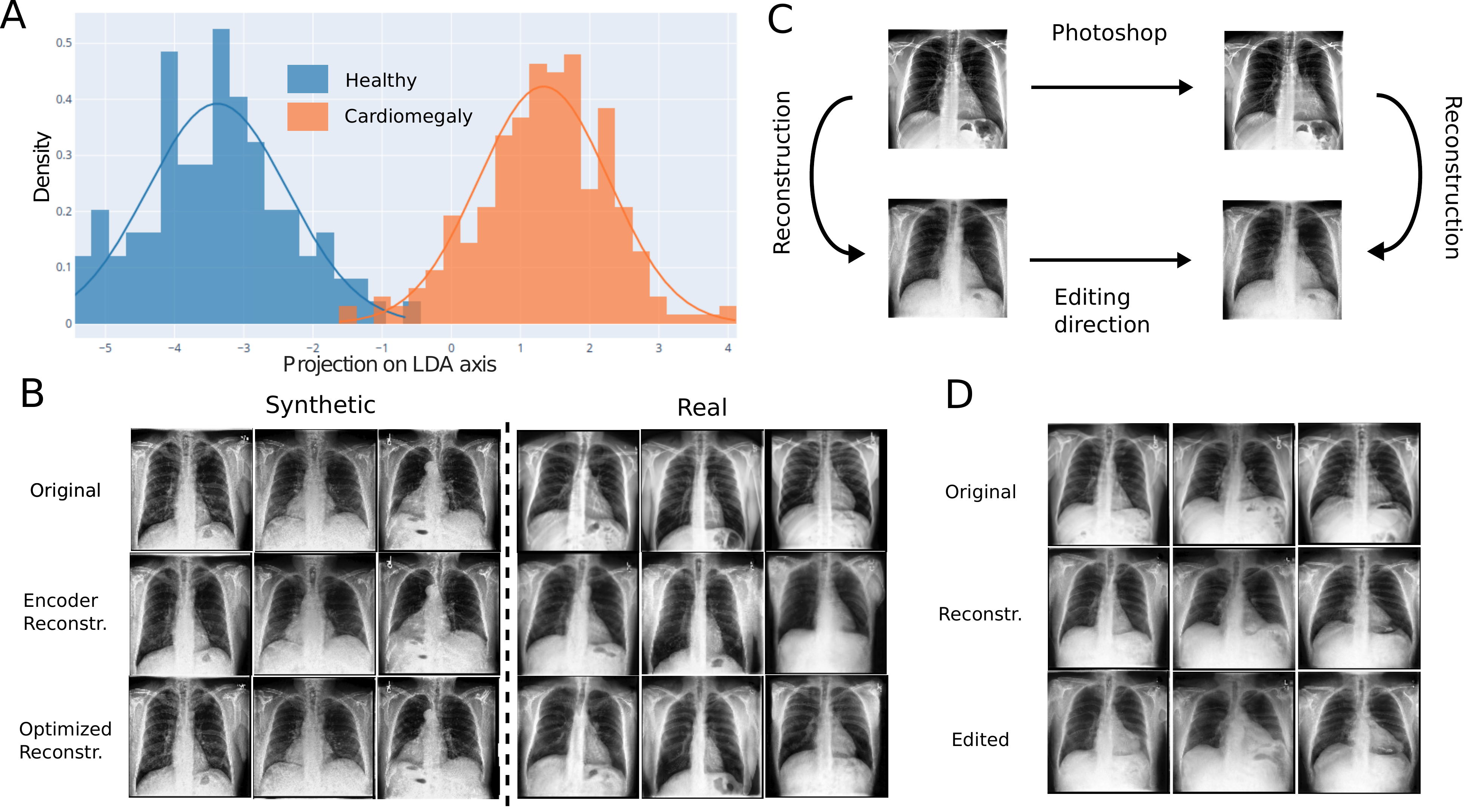}
      \caption{(A) LDA results; (B) Examples of generated and real images as well as their reconstructions; (C) Scheme for calculating a "direction of cardiomegaly" in latent space; (D) Image edits.}
    \label{fig:ExpResults}
    
\end{figure}

In this section, we gauge to what extent $G$ and $E$ fulfill conditions (i) - (iii).  In Fig.~\ref{fig:ExpResults}B, we show synthetic images generated by $G$ (first row, left). We find, $G$ is capable of producing realistic looking chest X-rays $x$. As for $E$'s capability of inverting $G$, we show corresponding reconstructions $G(E(x))$ (second row, left). While some minor differences remain, resulting images correspond to source images to a large extent. For comparison, we show reconstructions obtained by directly optimizing $\hat{w}$ s.t. $\hat{x}=G(\hat{w})\approx x$ (last row, left). In this case, the resulting reconstructions are almost perfect. 

To assess the sufficiency of $G$, in Fig.~\ref{fig:ExpResults}B, we also consider real chest X-rays (first row, right) and present reconstructions via $E$ (second row, right) and via direct optimization in latent space (last row, right). 
Encoder reconstructions reproduce source images on the organ level. More importantly, we find that reconstructions from optimized $\hat{w}$-s agree with source images to a high degree, suggesting that $G$ is indeed sufficient. Furthermore, $E$ is found to fulfil condition (iii) at least to certain approximation. 

To address characteristic (ii), we considered two classes of real chest X-rays, one being healthy patients ($n_1$=$100$ images), the other being patients with the medical condition of cardiomegaly ($n_2$=$250$ images), i.e. an abnormal enlargement of the heart. For all images, we obtained corresponding latent codes by optimization in latent space. Employing linear discriminant analysis (LDA), we investigated whether the resulting two sets of codes are linearly separable. Projections of all images along the optimal LDA-direction are displayed in Fig.~\ref{fig:ExpResults}A. This procedure resulted in two Gaussian distributions, the mean values of which are separated by about $4.8\times$ their mean standard deviation. This implies that the two classes are linearly separable in $W$,  suggesting that $G$ indeed exhibits at least partial disentanglement of medical conditions in latent space, i.e. fulfils condition (ii). 

Lastly, we asked whether we could use $G$ and $E$ to implement disease-aware image editing. To this end, we chose a random real chest X-ray $I$ from a healthy individual. By means of direct manipulation of the image in Adobe Photoshop, we created a modified version $I^*$ that resembles a chest X-ray of the same person, albeit with an enlarged heart (see Fig.~\ref{fig:ExpResults}C). Using $E$, we determined a direction $v$ in latent space by normalization of  $E(I^*)-E(I)$. The resulting vector $v$ is conceptualized as pointing towards the "direction of cardiomegaly", whilst keeping all other person-specific characteristics unchanged. Indeed, when translating optimized latent codes for randomly chosen real chest X-rays in the directions of $v$, we find that images generated from the translated codes exhibit an enlarged heart (see Fig.~\ref{fig:ExpResults}D, last row) while preserving their identity. These results demonstrate that we are indeed able to transfer a specific disease to healthy chest X-rays by means of our generator-encoder system. 

\section{Conclusion}
In this work, we present a proof of concept for the disease-aware editing of chest X-rays at the organ scale. To this end, we present a generator-encoder system realizing the characteristics (i)-(iii) at least to a limited degree. In particular, our preliminary empirical results show that we can generate realistic synthetic chest X-rays and edit real ones in a disease-aware manner by computing a latent direction using our encoder. Here we focus on image edits at the organ scale for cardiomegaly. The proposed method, however, has the potential to work also for other health conditions. Most likely, this will necessitate improvements such as increasing the resolution of generated images and improving the efficiency of encoder embeddings. The relevant code for this work can be accessed at \url{http://covbase.igib.res.in/}.

\section{Acknowledgement}
All the authors duly acknowledge CSIR-Institute for Genomics and Integrative Biology, CSIR-Central Electronics Engineering Research Institute,Caring Research and the CovBase initiative for providing computational and intellectual support.

\section{Broader Impact}
In this work, which originated as an attempt to build a COVID-19 classifier, we investigate a new paradigm with respect to image generation, reconstruction, and disease-aware editing in the context of chest X-rays. 
While the development of image-manipulation tools warrants concerns with respect to potential threats to the integrity of sensitive patient data, we believe a number of beneficial applications of the proposed approach outweigh its potential risks.
The generation of synthetic chest X-rays mimicking a given disease type is a viable strategy for alleviating the aforementioned scarcity of labelled image data in this domain. 
As such it is likely to boost the accuracy of automated diagnostic tools.
Disentangling disease-related from healthy image features holds the potential for building Explainable-AI systems, i.e. white-box algorithms highlighting image features indicative of disease, which can be leveraged for augmenting the accuracy and efficiency of pathologists \cite{TosPul20}. 
Beyond the scope of the medical domain, these developments will likely also have repercussions for the broader field of GAN-research and distribution learning in general.

\bibliographystyle{plain}
\bibliography{bibliography}

\end{document}